\documentstyle[11pt,epsfig]{article}
\author {M. Neek-Amal$^1$~  and  F. M. Peeters$^2$ \\
\small $^1$ Department of Physics, Shahid Rajaee Teacher Training
University,
Lavizan, Tehran 16788, Iran.\\
 {\small $^2$Departement Fysica, Universiteit
Antwerpen, Groenenborgerlaan 171, B-2020 Antwerpen,
 Belgium. }}
\begin{document}
\title{\bf Buckled circular monolayer graphene: a graphene nano-bowl}
\date{\today}
\maketitle
\begin{abstract}
 We investigate the stability of circular monolayer graphene subjected
to a radial load using non-equilibrium molecular dynamics
simulations. When monolayer graphene is radially stressed, after
some small circular strain ($\sim 0.4\%$) it buckles and bends into
a new bowl like shape. Young's modulus is calculated from the linear
relation between stress and strain before the buckling threshold,
which is in agreement with experimental results. The prediction of
elasticity theory for the buckling threshold of a radially stressed
plate is presented and its results are compared to the one of our
 atomistic simulation. Jarzynski equality is used to
estimate the difference between the free energy of the
non-compressed states and the buckled states. From a calculation of
the free energy we obtain the optimum radius for which the system
feels the minimum
boundary stress. \\


\end{abstract}
\maketitle

\section{Introduction}
Graphene, is a newly discovered almost flat one-atom-thick layer of
carbon atoms which exhibits  unique electronic properties and
unusual mechanical properties~\cite{lee,Giem2008}. At non-zero
temperature graphene is not a perfect flat sheet. Recent
experimental observations have found ripples in suspended sheets of
 graphene~\cite{mayer}. Atomistic simulations of graphene suggest
that the strong bonds between carbon atoms in graphene is
responsible for the ripples~\cite{fasolinonature}.  Controlling the
creation of one and two-dimensional periodic ripples is possible in
suspended graphene sheets by using both externally and thermally
generated strains~\cite{naturenanotechnology}.

 Tensional strain in monolayer
graphene affects the electronic structure of graphene. For example
strains larger than $15\%$ changes graphene's band structure and
leads to the opening of an electronic gap~\cite{naturephys}. The
strain can generate a bulk spectral gap in the absence of
electron-electron interaction as was found within linear elasticity
theory in combination with a tight-binding
approach~\cite{castroneto}. In recent experiments the buckling
strain of a graphene sheet that was positioned on top of a substrate
was found to be six orders of magnitude larger (i.e. $0.5-0.6\%$)
than for graphene suspended in air~\cite{arxiv2010}. Furthermore,
some experiments showed that compressed rectangular monolayer
graphene on a plastic beam with size 30$\times$100 $\mu m^2$ is
buckled at about 0.7$\%$ strain~\cite{compressionamall}. These
experiments indicate that in spite of the infinitely small thickness
of the monolayer of graphene, supporting graphene by various
substrates enhances the buckling strain considerably beyond the very
small values ($\sim 10^{-6}\%$) obtained from elasticity theory (for
very thin plate) using classical Euler analysis~\cite{arxiv2010}.
Although elasticity theory predicts that the very small thickness of
graphene yields a zero flexural rigidity but the bond-angle effects
on the C-C interatomic interaction (three body terms in Brenner's
potential) gives a non zero flexural rigidity for
graphene~\cite{bondstrength}. Furthermore, elasticity theory
indicates that the dimensionless critical load (buckling load)
parameter depends on the thickness, radius and mechanical properties
of the circular plate. It is found that it decreases nonlinearly
with the thickness-radius ratio and approaches a constant value for
very small thickness-radius ratios, i.e. the Kirchhoff buckling load
factor, independent of the boundary
conditions~\cite{bookbuckl,sakhaeepour,buck1,buck2}.

Using computer simulations it is possible to calculate various
thermomechanical properties of new nano scale objects and to propose
 new nano-structures~\cite{yakobson}. Graphene as a nano balloon~\cite{scott2}
 is such an example where, first-principles
density functional theory was used to investigate the penetration of
helium atoms through a graphene monolayer with
defects~\cite{peeters}. Density functional theory has been employed
to investigate the structural, electronic and mechanical properties
of rectangular graphene nanoribbons subjected to boundary
stress~\cite{JPCMreferee} and central load~\cite{nanolettreferee}.
Larger values for the 2-dimensional Young's modulus with respect to
the one for bulk graphene ~\cite{nanolettreferee} were predicted and
a decreasing  2-dimensional Young's modulus with respect to the size
was reported for small size ($\sim$ 1-5 nm) of graphene
nanoribbons~\cite{JPCMreferee}. Recently, we studied the possible
deformations of rectangular graphene nano-ribbons (GNRs) at room
temperature under axial strain for both free and laterally supported
boundary conditions~\cite{neekamalsubmitted}. We found several
longitudinal sinusoidal deformation modes for different sizes of
GNRs and investigated the thermal stability of buckled GNRs. In such
a study the calculation of the free energy difference is  important
in order to compare the stability of various structural states of
stressed graphene. Thermodynamic integration and the perturbation
method are two time consuming methods that have been used to
calculate the difference in free energy~\cite{hui}. But a more
approporiate approach is based on the Jarzynski~\cite{jar} identity
that is valid for  non-equilibrium simulations and is applicable to
the study of the stability of compressed and stretched graphene.

Here, we study the mechanical stiffness of circular monolayer
graphene (CMG) by investigating its linear response to strain. Using
the stress-strain curve in the linear regime, we obtain Young's
modulus and the pre-stress in the CMGs which we found to be in good
agreement with previous experimental and theoretical studies. Our
simulations introduce a way to calculate pre-stresses which, to the
best of our knowledge, is the first time that they have been
calculated. After continuing the compression, CMG starts to buckle.
The buckling strain in most cases are larger than the  predictions
from elasticity theory. Moreover the stability of a new state of
radial compressed CMG which we call \emph{nano-bowl} graphene (NBG),
and the optimum size of CMG are studied by using atomistic
simulations and the  Jarzynski identity for the calculation of the
free energy~\cite{jar}. To the best of our knowledge this is the
first time that such a free energy calculation approach is used in
the field of thermomechanical properties of graphene. Recently,
Colonna \emph{et al} used the free energy integration based method
to explain the melting properties of graphite
~\cite{freeenergyfasolino}. Our nano-scale bowl like structure for
CMG is a buckled state of the thinnest material in the world.

This paper is organized as follows. In Sec.\,2  we  introduce the
atomistic model, the simulation method and a  discussion on the
Jarzynski's equality as a method to estimate the free energy
difference between two states within a non-equilibrium simulation.
In Sec.\,3 we present our numerical results. Estimation of the
Young's modulus and pre-stresses, estimation of the buckling
thresholds, the free energy changes during compression of the CMG
and the stability of the obtained bowl like shape of the compressed
CMG are the main results of this study. After giving our results for
buckling, we present the prediction from elasticity theory for the
buckling of a circular plate and compare them to the one obtained
from molecular dynamics simulation. Also a discussion on the effect
of the thickness of graphene on its mechanical properties is
presented. In Sec.\,4 we conclude the paper.

\section{Method and model}
\begin{figure}
\begin{center}
\includegraphics[width=1.0\linewidth]{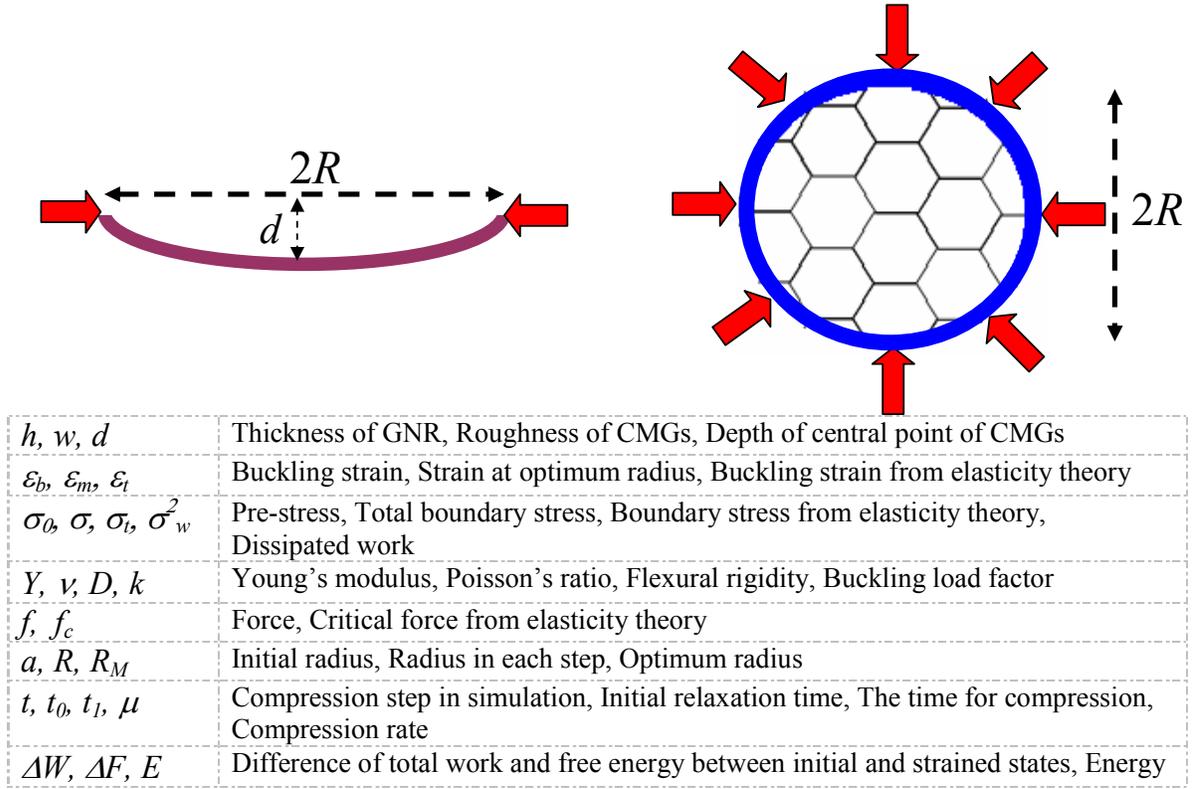}
\caption{(Color online) Schematic model for a circular plate under
radial stress with clamped boundary condition. Left picture is the
side view and the right picture is the top view. We list all
relevant variables describing the CMGs. Arrows show the applied
boundary forces.} \label{fig0}
\end{center}
\end{figure}

 Classical atomistic molecular dynamics
simulation (MD) is employed  to simulate compressed suspended CMG
using Brenner's bond-order potential~\cite{brenner}. A rigidly
clamped boundary condition was imposed on a CMG with inner initial
radius $a$. The boundary atoms were located in the radial distances
 $[a,a+2~\AA]$. In Fig.~\ref{fig0} we depict a
schematic model for both side view and top view of a CMG under
radial stress with clamped boundary condition and list all relevant
variables describing CMG under radial stress. We simulated the
system at room temperature and 50~K by employing a Nos'e-Hoover
thermostat. Total number of atoms inside the boundary varies from
2224 to 18148
 and there are 300 to 850 atoms at the
boundaries. Initially the coordinates of all atoms are put in a flat
surface of a honey comb lattice with nearest neighbor distance equal
to 1.42~\AA~ and the initial velocities in each direction were
extracted from a Maxwell-Boltzman distribution for the given
temperature.

First we equilibrated the system during a time $t_0$ (\emph{process
`S'}), then we started to compress the system  in the radial
direction (\emph{process `I'}) with the rate of compression
$\mu$=0.48~m/s. The radius ($R$) at time $t$ during the process `I'
and the absolute value of the boundary strain are
\begin{equation} R=a-\mu
(t-t_0)~;~\epsilon=\frac{\mu (t-t_0)}{a}.\label{strain}
\end{equation}
When compressing the system, the radius of the system is reduced by
very small steps (0.012~\AA) to achieve a quasi-stable situation.
After each compression step,  we wait 2.5~ps to allow the system to
relax. Note that the initial radius $a$ is for a perfect flat sheet
which is not necessarily the equilibrium radius of the system (i.e.
optimum size; $R_M$) as we will find later.
 At the end of process `I' we relaxed the system to the new
obtained radius $R_{comp}=a-\delta$ where $\delta=\mu(t_1-t_0)$,
(\emph{process `II'}). For example for $a=10$ nm, setting
$t_1$=500~ps with $t_0$=250~ps gives a small radial strain
$\epsilon=$1.2$\%$
\\

An important theoretical property is the free energy which is a
global property which depends on the extent of the configuration
space accessible to the system. In general this phase space is huge
, but in practice it is sufficient to obtain an estimate of only the
difference of the free energy between two related states of the
system. This difference corresponds to the relative probability of
finding the system in one state as opposed to the other.
Thermodynamic integration and perturbation method are two common
approaches to calculate this difference in the free
energy~\cite{freeenergyfasolino}. In traditional methods (for
example in the perturbation method) only when two states are close
in their phase-space coverage, the convergence of the algorithm is
good~\cite{hui}.  A robust and more general approach, which is also
valid for non-equilibrium thermodynamics, is based on the Jarzynski
equality~\cite{jar} whose
 only requirement is a slow (quasi static) transformation between
 the two states (which not necessarily have to be close in phase
 space).

Due to the application of external forces on the boundary of the
CMGs, an equilibrium approach is no longer applicable and
non-equilibrium molecular dynamics simulation is needed. In the
present paper we are interested  to find the change in the
difference between the free energy of two states: the initial
non-compressed graphene membrane and a new found state of compressed
CMG which we call \emph{nano-bowl} graphene (NBG).  In the ideal
case with infinitely small or quasi-static transformation of the
parameters of the system, the system will evolve along a path
starting from state `A' to state `B'. In such a case the total work
done on the system is equal to the difference between the Helmholtz
free energy of the two states independent of the used path (i.e.
second law of thermodynamics). By contrast when the system evolves
with a \emph{finite} rate, as in common molecular dynamics
simulations, the total work depends on the microscopic initial
conditions and the above mentioned equality fails, i.e, $\Delta
F=F_B-F_A\neq W$.

 Independent of the path and the evolution rate between two thermodynamical states,
  Jarzynski found the following equality  between the difference
of the free energy and the total work done on the system~\cite{jar}
\begin{equation}
\Delta F=-\beta^{-1}\ln\langle{\exp(-\beta W)}\rangle,\label{jeq}
\end{equation}
where $\beta=1/k_BT$. The averaging is done over the possible
realizations of the switching process between the initial and the
final state. Notice that Eq.~(\ref{jeq}) connects the difference of
the equilibrium free energy between states `A' and `B'  and the
non-equilibrium work done when going from  `A' to `B' even in the
presence of a thermostat~\cite{jar,statjar}.

For sufficiently slow switching between the states `A' and `B' the
obtained values for $W$ are distributed Gaussian and only the two
first terms of the expansion according to Eq.~(\ref{jeq}) will
survive, i.e. $\Delta F=\langle W\rangle-\frac{\beta
{\sigma_w}^2}{2}$, here the second term is the dissipated work as a
consequence of  fluctuations in $W$~\cite{jar}. When the parameters
are changed with  a finite rate, $W$ will depend on the microscopic
initial conditions of the system and the reservoir (i.e. thermostat)
and the average of $W$ becomes larger than $\Delta F$,
\begin{equation}
\langle W \rangle \geq \Delta F.\label{Wpositive}
\end{equation}
Higher evolution rates result in a larger difference $\langle W
\rangle -\Delta F$ (larger $\sigma_w$) (for more discussion see
Ref.~\cite{hui}). Moreover it is important to mention that the value
of $exp(-\beta W)$ in Eq.~(\ref{jeq}) should be obtained through a
high precision calculation.

\section{Simulation results}
\subsection{Young's modulus and pre-stress}

 We compressed the system during process `I' and calculated the
total force applied on the boundary atoms  at time $t$ ($t_0< t<
t_1$) with radius $R$. We find the applied force on the inside
boundary atoms ($f$)  as the minus of the force acting on the
boundary atoms. The corresponding radial boundary stress on the
inside region is calculated by
\begin{equation}
\sigma=\frac{f}{2\pi Rh}.
\end{equation}
We performed several simulations for different radius $a$ at
$T$=300~K, $t_0$=25~ps, and the thickness of CMG $h\simeq0.335$
nm~\cite{lee}.
 Figure~\ref{figFt}(a) shows the variation of the radial boundary
stress (-$\sigma$) versus radial strain, i.e. stress-strain
 curve, for the two radii $a$=6 and 12~nm.
We found that before buckling, during process `I',  the radial
boundary stress increases linearly, and satisfies a simple linear
relation between stress and strain, $\sigma=\sigma_0+ \epsilon Y$.
For small strains  anharmonicity is weak~\cite{poisson}.   By
calculating the slopes of the dashed lines given in
Fig.~\ref{figFt}(a), we found the (effective) Young's modulus ($Y$)
for different radii and listed them in Table~\ref{table0}. The
obtained Young's modulus are in the range of recent experimental
measured values, i.e. 0.5-1.01 TPa~\cite{lee,Frank}.
 Notice that the obtained Young's modulus depends on the used value for the thickness.
In the case of a single atom thick membrane (graphene), one should
be careful about the definition and dimension of the Young's
modulus. Since external stress is applied on the boundary of
graphene which is an  almost one dimensional curve, (here the
perimeter of the CMG), thus the exact unit for $Y$ should be N/m (or
TPa\AA).~For comparison purposes we listed $Y$s with TPa
\AA~unit~\cite{JPCMreferee} in the second row of Table~\ref{table0}
and also showed it in the inset of Fig.~\ref{figFt}(a). However
usually by dividing this number by the thickness of graphene, $Y$ is
reported in Pascal~\cite{lee}. Here we used for the thickness
$h$=3.35~\AA~as done in the experimental papers and therefore there
should be no misinterpretation. In Fig.~\ref{figFt}(a) the
y-intercept, $\sigma_0$, is the pre-stress in the system, which
occurs because our initial system with radius $a$ is not in an
equilibrium state with optimum radius. The pre-stresses are 2.34 and
1.946 GPa for the two radii $a$=6, 12~nm, respectively. These
pre-stresses are of the same order of magnitude as the values, e.g.
2.23 and 1.19 GPa that were measured on two flakes with diameter 1
and 1.5 $\mu m$~\cite{lee} and are also  in agreement with those
found in our previous study on GNRs~\cite{neekamalsubmitted}. At
zero strain, the sign of the pre-stresses indicate that the inside
atoms exert an inward stress on the boundary atoms. Furthermore,
note that here (except for the two smallest CMGs) Young's modulus
varies (around 0.525 TPa) with respect to the size of the CMGs.
Young's modulus of graphene nano-ribbons depends on the size of the
system. The chrematistic length of graphene (8-10
nm~~\cite{fasolinonature}) is a measure of the range over which the
deformations in one region of graphene are correlated with those in
another region, so the applied boundary stresses on the edges do not
affect the system beyond this characteristic length. We expect that
our system with radius less than the characteristic length ($2a<$12
nm) will show a larger $Y$. But we believe that this behavior
depends also  on the applied boundary condition, i.e. clamped
boundary condition. Faccio \emph{et al}~~\cite{JPCMreferee} and
Hod~\emph{et al}~~\cite{nanolettreferee} used density functional
theory and found a larger  value for $Y$ for smaller size graphene
nano-ribbons (less than 6 nm) and reported a decreasing  $Y$ with
increasing size of nano-ribbons.

\subsection{Buckling results}
\subsubsection{Molecular dynamics predictions for buckling of CMG}
\begin{figure*}
\begin{center}
\includegraphics[width=0.32\linewidth]{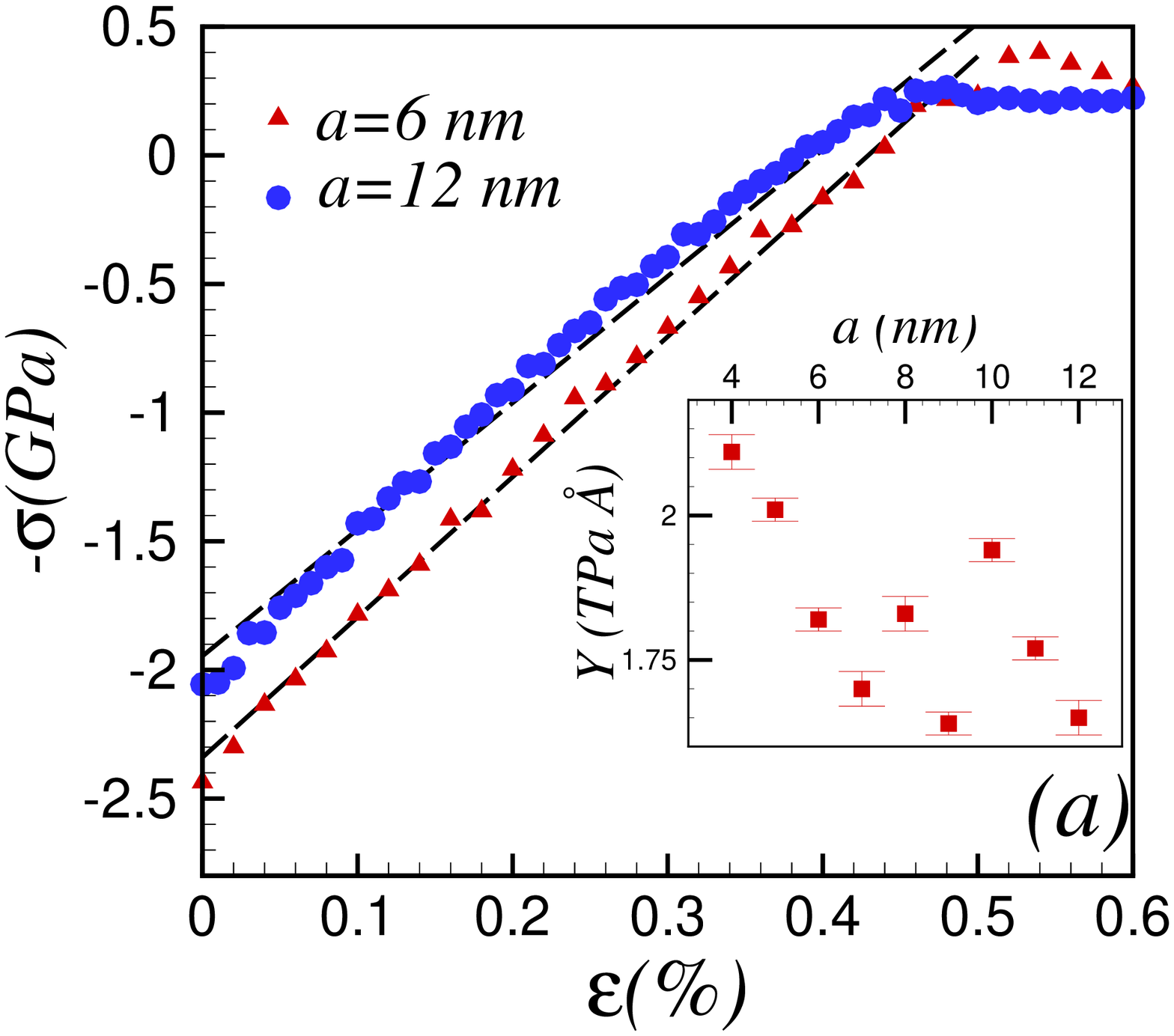}
\includegraphics[width=0.32\linewidth]{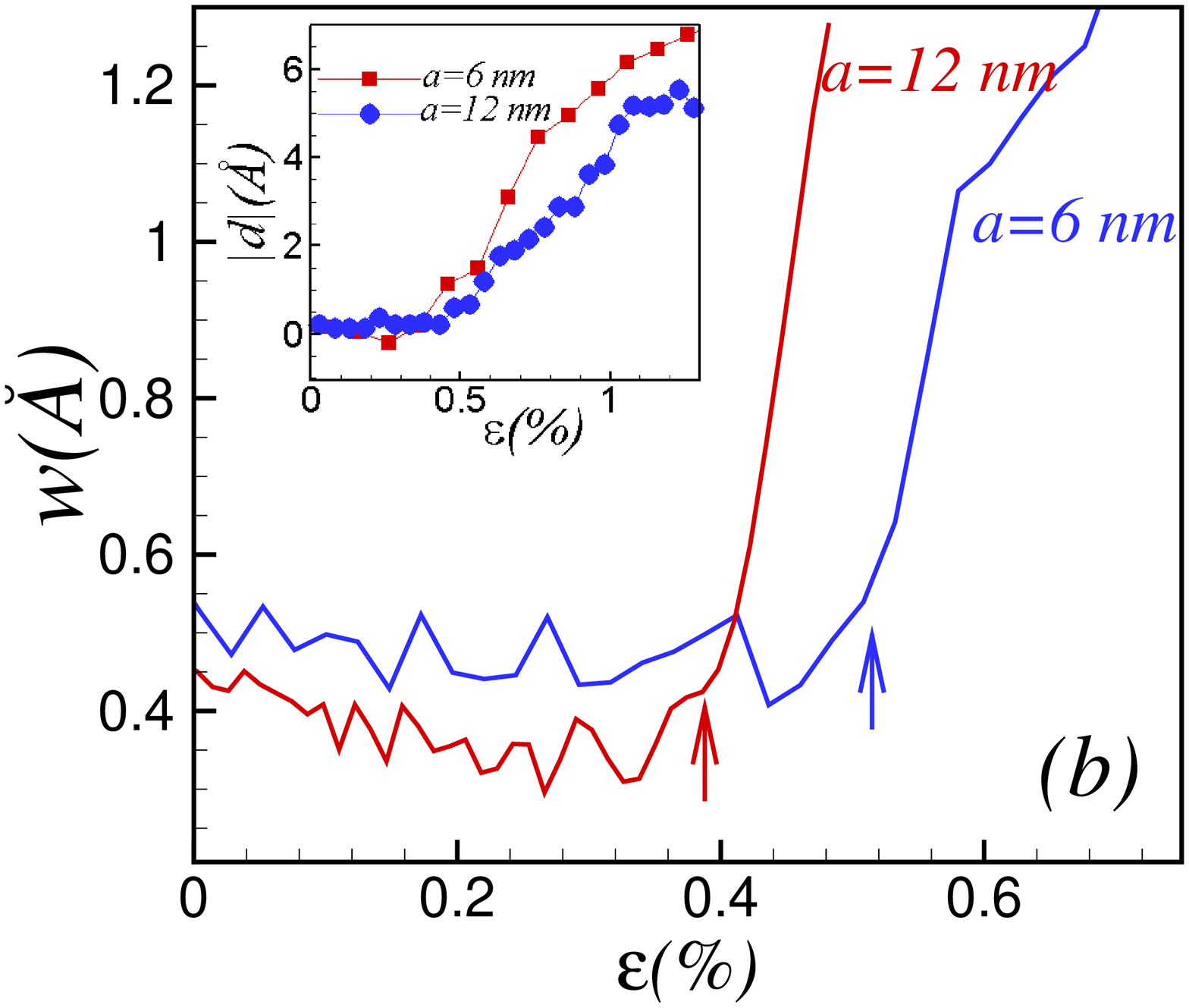}
\includegraphics[width=0.32\linewidth]{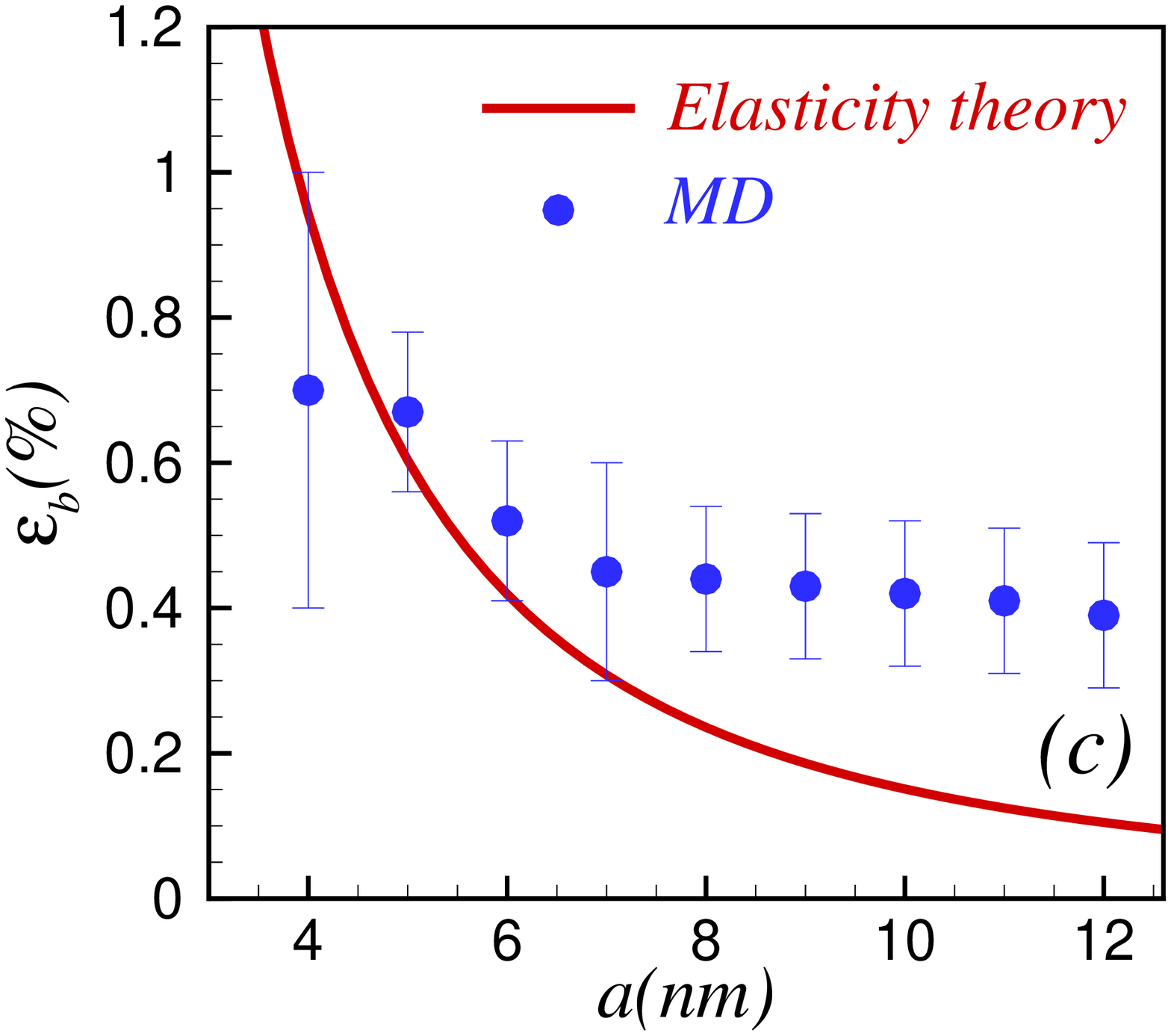}
\caption{(Color online) (a) Variation of the radial
 in-plane stress (radially inward direction) applied on
the boundary versus the radial boundary strain for two different
radii of circular monolayer graphene, i.e. 6 and 12~nm. The dashed
lines are fitted to the numerical results in the region before the
buckling threshold. The inset shows the Young's modulus as a
function of the radius.~(b) Variation of the root mean square
displacement of the z-component of atoms of CMGs versus the applied
boundary strain for two radii $a$=6 and 12 nm. The inset shows the
variation of the depth of the buckled circular monolayer graphene
versus strain. (c) Circular dots are from our numerical simulations
for the buckling strain and the solid line  is the prediction from
elasticity theory when  the thickness of graphene was set to 0.335
nm.} \label{figFt}
\end{center}
\end{figure*}

Beyond the linear regime and during process `I', we estimate the
critical buckling stress. The rapid increase of the lateral
deflection with load near the buckling threshold defines
experimentally the buckling stress~\cite{bookbuckl}. In our
simulations we estimated the buckling points as the points where the
roughness of the system suddenly increases. In Fig.~\ref{figFt}(b)
we show  the variation of the root mean square displacement -$w$-
(rms) of the z-component of the CMG versus applied strain for two
radii $a$=6 and 12~nm. The
 buckling point is obtained with a  uncertainly which increases with decreasing size of
 the membrane which results in larger error bars on the buckling strain for small systems.
The sudden change in the slope of these curves are directly related
to the buckling threshold (solid arrows in Fig.~\ref{figFt}(b)). For
small applied strain (i.e. below the buckling points) $w$ exhibits
random fluctuations which are due to the applied boundary stress and
the thermal fluctuations. The inset of Fig.~\ref{figFt}(b) shows the
depth, $d$ (the z-components of the atoms in the center of CMG), of
the buckled CMG versus strain which exhibits a clear nonlinear
dependence.  It is interesting to note that we found that for larger
sizes (i.e. $a \geq 8$ nm) the buckling strain is only slightly
modified. In this limit the size of CMGs are beyond the
characteristic length of graphene~~\cite{fasolinonature}. Usually,
the mechanical properties of graphene are found to be unchanged for
systems that have larger sizes than the characteristic
length~\cite{bulkgraphene}. Therefore it is not necessary to
simulate systems with larger radii.

 The boundary strains ($\epsilon_b$) are
listed in Table~\ref{table0}. The obtained values  show a non-linear
decrease with increasing radius as shown by solid dots in
Fig.~\ref{figFt}(c). Larger buckling strains are found for smaller
systems. This behavior is confirmed by recent  quantum molecular
dynamics simulation~\cite{physcaE}, where for a small system with
size 1.99$\times$0.738 $nm^{2}$ a buckling strain of 0.8$\%$ was
found.

\begin{table}
\begin{tabular}{|cccccc|}
\hline
 $a~(nm)$&4.0&5.0&6.0&7.0&8.0\\

  \hline
$Y$(TPa)&0.631$\pm$8.0&0.601$\pm$6.0&0.543$\pm$7.0&0.507$\pm$8.0&0.546$\pm$8.0\\
$Y$(TPa~\AA)&2.11$\pm$0.03&2.01$\pm$0.02&1.82$\pm$0.02&1.70$\pm$0.03&1.83$\pm$0.03\\
$\epsilon_b(\%)$&0.71$\pm$0.3&0.67$\pm$0.1&0.52$\pm$0.1&0.45$\pm$0.2&0.44$\pm$0.1\\
\hline
 $a~(nm)$&9.0&10.0&11.0&12.0& \\
\hline
$Y$(TPa)&0.490$\pm$9.0&0.579$\pm$8.0&$0.528\pm$8.0&$0.492\pm$9.0&\\

$Y$(TPa~\AA)&1.64$\pm$0.03&1.94$\pm$0.02&$1.77\pm$0.02&$1.65\pm$0.03&\\

$\epsilon_b(\%)$&0.43$\pm$0.1&0.42$\pm$0.1&0.41$\pm$0.1&0.39$\pm$0.1&\\
\hline
\end{tabular}
\centering \caption{Young's modulus and the buckling strain for
different radius of the circular monolayer graphene.}\label{table0}
\end{table}

\subsubsection{Elasticity theory predictions for a buckled circular
plate}

We will compare our results with predictions of elasticity theory
for the buckling of a circular plate. The sudden structural
deformation of a continuum membrane subjected to high compressive
stress is called buckling. At the point of failure the actual
compressive stress is less than the ultimate compressive stresses
that the material is capable of withstanding. The boundary condition
and the elasticity of the plate are two important parameters in
determining the buckling critical load. The buckling occurs when the
disturbing moment of a material equals the restoring
moment~\cite{bookbuckl}. The governing equation for a simple bar
with length $L$ can easily be obtained by considering the curvature,
bending moment and disturbing moment of the bar, i.e. Euler's column
theory~\cite{bookbuckl}, resulting $\frac{\pi^2 P}{L^2}$ for the
buckling load, where $P$ is a parameter which is related to Young's
modulus ($Y$) and the moment of inertia of the rod cross-sectional
axis that is perpendicular to the  buckling plane. For the case of
circular plate under radial compressive force, according to
continuum elasticity theory one can define the strain energy as the
restoring energy in the plate due to the external forces. The
minimization of the strain energy of a circular plate based on
Trefftz's theory for the linear and nonlinear components of the
strain and the stress tensors results into two coupled differential
equations~\cite{buck1,buck2}. Traditionally it is better to define a
dimensionless quantity -critical or buckling load factor, $k$ as
$k=\frac{f_c a^2}{D}$. Here $f_c$ is the critical radial compressive
force on the boundary per length, $D=Yh^3/(12(1-\nu^2))$ is the
flexural rigidity of the plate with thickness $h$, $\nu$ is the
Poisson's ratio  and $Y$ is the Young's modulus. Flexural rigidity
of a plate is defined as the force required to bend a rigid
structure to a unit curvature. Poisson's ratio is the ratio of
transverse contraction strain to longitudinal extension strain in
the direction of stretching force. In the case of axisymmetric
buckling and the assumption of no variation in the thickness,  the
solution of the equations yields the Kirchhoff buckling load factor,
$k_0(\simeq 14.6820)$, for different combinations of boundary
conditions. Different boundary condition for the circular plate
gives different buckling loads~\cite{buck1} while the atomistic
structure of the plate is not included directly in the theory. For
the clamped boundary condition, where the perimeter of the circular
plate is not allowed for any movement, and in the absence of any
pre-buckling deformation (in the system), the equation for $k$ leads
to ~$k=k_0/(1+\frac{(h/a)^2k_0}{5(1-\nu)})$~\cite{buck1,buck2}. Note
that elasticity theory predicts that when the thickness-radius ratio
is set to zero we find $k=k_0$ and consequently the buckling load is
$ \frac{k_0D}{a^2}$ which is similar to the formula for the buckling
load of a bar, i.e. $\frac{\pi^2 P}{L^2}$. The radial compressive
force for a plate with radius $a$ can be written as
\begin{equation}
f_c=\frac{k_0 D}{a^2+\frac{h^2k_0}{5(1-\nu)}}.
\end{equation}
Dividing $f_c$ by the thickness yields the relationship between
stress and strain  as
\begin{equation} \sigma_t=\frac{k_0
h^2}{12(1-\nu^2)(a^2+\frac{h^2k_0}{5(1-\nu)})}Y=\epsilon_t Y,
\end{equation}
where $\epsilon_t$ is the theoretical  buckling strain. Note that,
elasticity theory predicts an inverse square dependence on the
radius: $\epsilon_t\cong\frac{k_0}{12(1-\nu^2)}\frac{h^2}{a^2}$
 which is shown by the solid line in Fig.~\ref{figFt}(c)
 (where $h\simeq3.35$ \AA~and $\nu=0.3$ is the typical Poisson's
ratio~\cite{poisson}). According to elasticity theory, below the
solid line the system is in the non-buckled (rippled) state and
above this line it is buckled~\cite{soft}. A rippled state can be
understood as any thermodynamical state of graphene before buckling
and under non zero boundary stress. Our results (except for $a$=4
nm) are above the thresholds found from elasticity theory. The
reason is that the considerable and non-zero flexural rigidity of
graphene, which is a consequence of strong $sp^2$
bonds~\cite{bondstrength}, is responsible for those
 larger buckling thresholds as
compared to those found  from linear elasticity theory.

Note that both theoretical and MD results are sensitive to the used
thickness of monolayer graphene. Larger thickness leads to a larger
difference between MD results and elasticity theory.  For example,
by using a smaller thickness which is closer to the interatomic
distance in graphene .i.e. $\simeq0.1$ nm, predictions from
elasticity theory are reduced by one order of magnitude.  Recently,
for the buckling experiments of embedded graphene the same
disagreement was reported~\cite{arxiv2010}. In fact, the thickness
of graphene is an important parameter which causes several new
unusual mechanical properties of graphene, particularly large
stiffness with respect to other materials (physically the
interpretation of the stiffness and the rigidity of single layer
graphene strongly depends on the very thin thickness of graphene).
 This issue is important also for studying mechanical stiffness of
 other carbon allotropes and leads to different values for Young's modulus etc.
  Please note that the thickness of graphene depends on the used interatomic potential, loading method etc~\cite{thickness}.
 We followed the approach used by experimentalists and used  $h\simeq0.335$ nm which is the interlayer spacing of graphite~\cite{lee}.

In order to induce significant changes in graphene's band structure
one needs even larger strains ($>15\%$) than the one we obtained
from the buckling threshold~\cite{naturephys}. Thus, in the case of
compressive strain, the shape of the buckled CMGs will change
appreciably before the electronic band structure is
modified.\\

\subsection{Free energy}
\subsubsection{Free energy change during compression of CMG}

Here we will focus on the system with $a=10$~nm. Since a study of
the stability of graphene needs longer simulation time, we first
relaxed the system during $t_0$=250 ps (process `S'), and then
compressed the system during $t_1$=500 ps (with the compression rate
$\mu=$0.48 m/s, the applied strain rate is 0.0048/ps). At the end of
process `I' we relaxed the system to the new obtained radius
$R_{comp}=a-\delta$ where $\delta=\mu(t_1-t_0)$, i.e., process `II'.
Setting $t_1$=500~ps gives a small radial strain
$\epsilon_{r}=$1.2$\%$ and a reduction of the radius by
$\delta=$1.2~\AA. Here we discuss  process `I' and in the next
section we return  to  process `II'.

Fig.~\ref{figw}(a) shows the variation of the total work performed
on the system when changing the radius  from the end of the `S'
state  to  the `II' state. Dashed curves are the results obtained by
performing simulations with different initial conditions. The solid
curves are the average $\langle W\rangle$. We found that by using a
small rate ($\mu$), 15 members of the ensemble are sufficient to
realize convergence of the exponential ensemble averages in
Eq.~(\ref{jeq}) (the reason is that the differences in $W$ between
each trajectory are small). Almost the same $W$ curves indicate that
the variance of the work distribution is small (here typically $
\sigma_w < 0.05$~eV). For a larger rate, a larger variance will be
found~\cite{neekamalsubmitted}.

In Fig.~\ref{figw}(b), by using the Jarzynski equality, we
calculated the difference of the free energy between the states with
radius $R$ and $a$, i.e. $F-F_S$ . In this curve $t$ is the
compression time which is related to the strain from
Eq.~(\ref{strain}). Each symbol (at time $t$) is related to a
quasi-static state during process `I'. Comparing $\langle W\rangle$
(Fig.~\ref{figw}(a)) and $F-F_S$ (Fig.~\ref{figw}(b)) shows that the
difference in free energy is smaller than the total work. However
the difference is small enough (very small $\sigma_w$ in
Table~\ref{table1}) such that the chosen number of different
simulations (i.e. 15) and the rate ($\mu$) are sufficient. The first
minimum in the process `I' is deeper for temperature $T$=300~K as
compared to $T$=50~K and occurred before the buckling threshold. We
call this minimum the `M' point. The radius of the system at state
`M', i.e. $R_M$,  is the optimum radius of the CMG where there is no
boundary stress. Considering the difference in the free energy for
the points `M', one can estimate $\Delta F=F_M-F_S$. Some of the
results for the difference in free energy, variance in $W$
distributions and $R_M$ are explicitly listed in Table \ref{table1}.
The free energy of point `S' is higher than those of point `M'. This
general behavior tells us that the starting states `S' are not the
relaxed states and as we showed in section 3.1 there is pre-stress
in the system.  From the optimum radius  reported in Table
\ref{table1}, we see that the radius of the `M' point for lower
temperature is closer to the radius of the `S' point, i.e.
${R_M}{(300~K)}<{R_M}{(50~K)}$.  The reason is that the equilibrium
state at lower temperature has a smaller out of plane deviation, or
simply its state is closer to a flat plane. Furthermore, note that
the conditions $R_M>R_b$ and $\epsilon_M=\frac{0.3\delta}{a}\simeq
\frac{\sigma_0}{Y}$ are always satisfied where $R_b$ is the radius
of the CMG at the buckling point.

Moreover, we calculated the ensemble average of the total energy per
atom, containing both kinetic and potential energy.
Fig.~\ref{figw}(c) shows the variation of the total energy per atom
($E/N$) versus time during the evolution  S$\rightarrow$II. As one
expects increasing the temperature decreases the absolute value of
the total energy because of the increasing  kinetic energy. In
contrast to the free energy and the total performed work on the
system, (Figs.~\ref{figw}(a,b)), here we have   non zero energy
values during processes `S' and `II'. Since the total energy of the
system increases for the evolution  S$\rightarrow$M, we conclude
that this evolution is an entropy dominated evolution ($\Delta
F-\Delta E <0$ as we see from Table~\ref{table1}). In contrast to
the total work done on the system, the average of the total energy
(Fig.~\ref{figw}(c)) has an overall different behavior (particularly
at the `M' points) from the curves shown in Figs.~\ref{figw}(a,b).
The increasing change (not decreasing change) in the slope of the
energy curves during process `I', point to the `M' points. We
conclude that the equilibrium relaxed size of the simulated system
could not be obtained by looking only at the total energy or the
potential energy and that free energy considerations are necessary.

\subsubsection{Stability of the nano-bowl}
\begin{figure*}
\begin{center}
\includegraphics[width=0.32\linewidth]{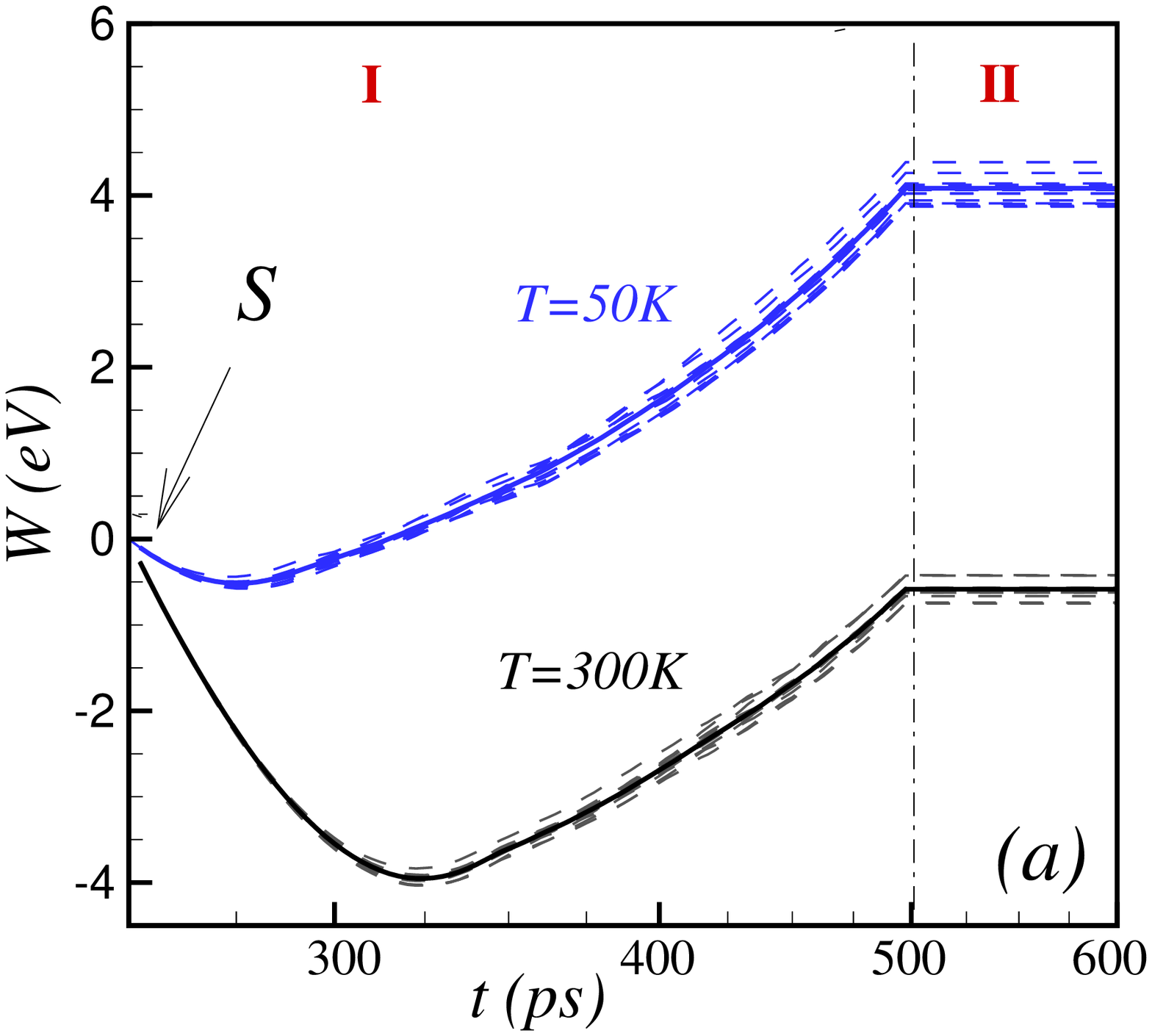}
\includegraphics[width=0.32\linewidth]{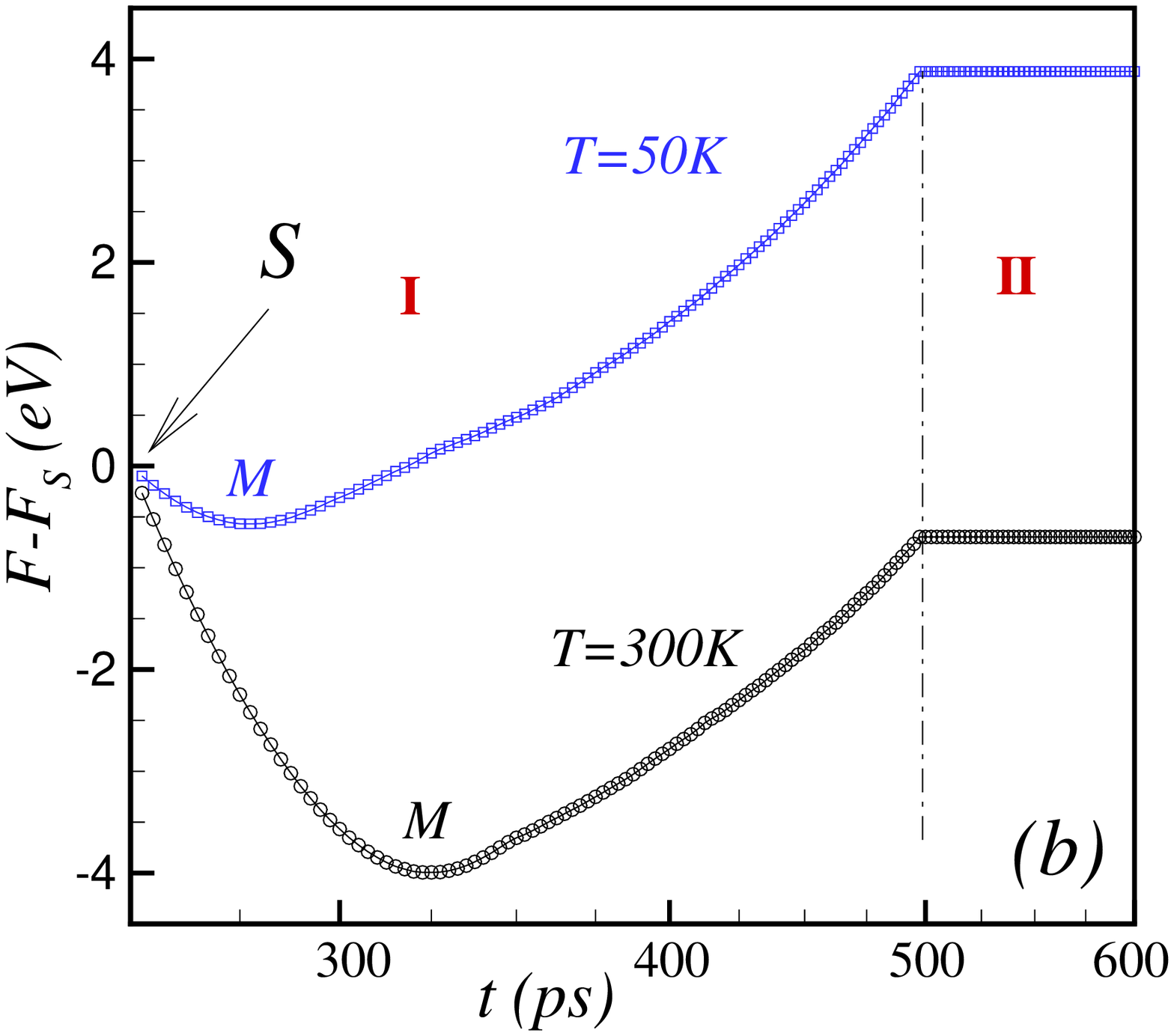}
\includegraphics[width=0.32\linewidth]{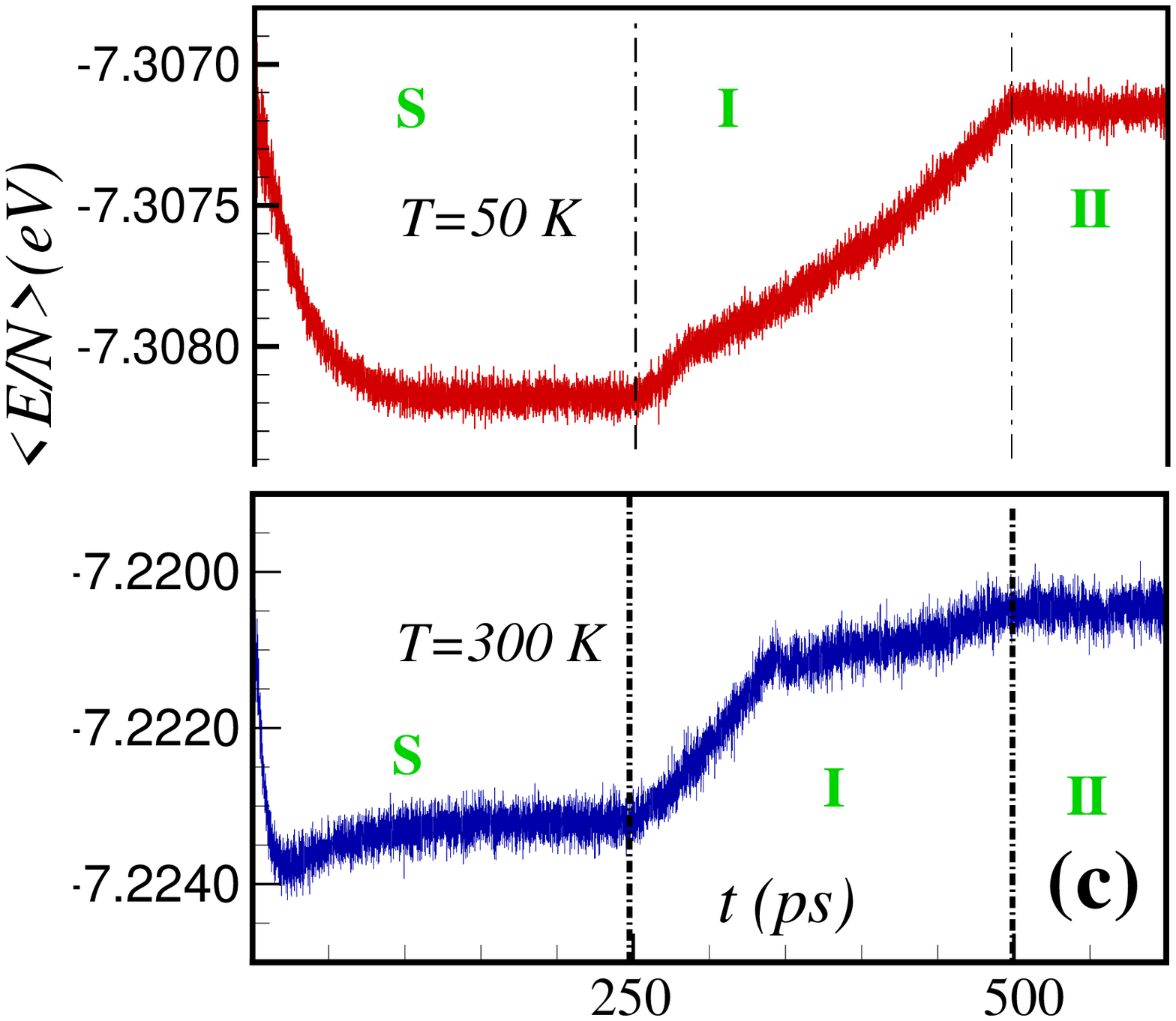}
\caption{(Color online) (a) Total work done on the system (with
initial radius a=10\,nm and two different temperatures, i.e. 50 K
and 300 K) during the compression of  a circular monolayer graphene
where the  total boundary strain is $\epsilon_{r}=1.2\%$. During
process `I' the system is compressed for 250~ps and it reaches the
radius $a$-$\delta$ at the end of this process. During the process
`II' we allow the system to relax in the compressed state for
250~ps. Solid curves are the average of the dashed curves. (b)
Corresponding difference in the free energy (using Jarzynski
equality). (c) Ensemble average of the total energy per atom of
circular monolayer graphene during evolution S$\rightarrow$ II.
\label{figw} }
\end{center}
\end{figure*}

\begin{figure}
\begin{center}
\includegraphics[width=0.8\linewidth]{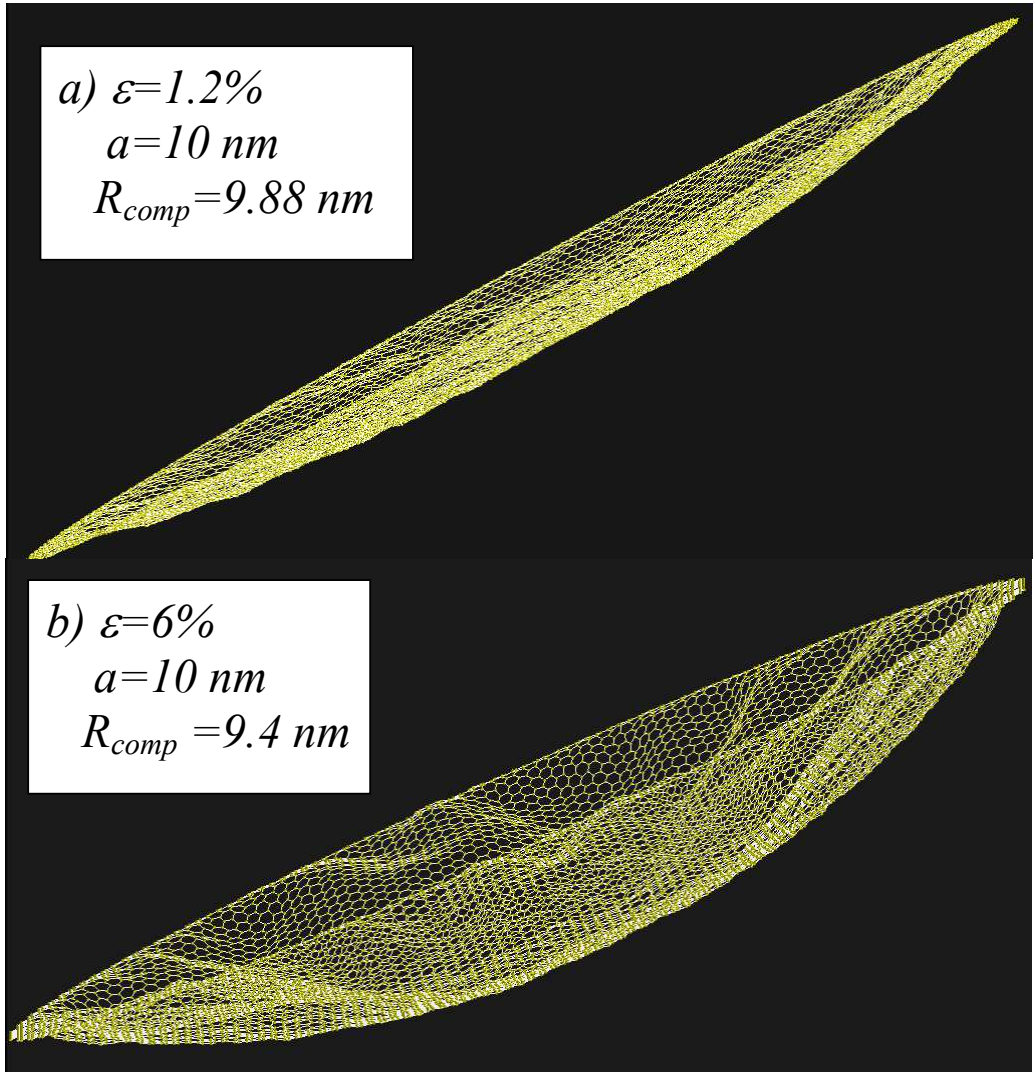}
\caption{(Color online) Two snap shots of buckled circular monolayer
graphene after applying radial boundary strains: (a)
$\epsilon_{r}=1.2\%$ and (b) $\epsilon_{r}=6\%$~ with initial radius
$a=10$~nm~at $T$=300~K. \label{figbowl} }
\end{center}
\end{figure}

 After compressing, CMG reaches the process `II', where the radius of the CMG
  is $R_{comp}$ and we arrive at the state NBG. This new state
is observed after the buckling threshold. Therefore all states after
the buckling threshold are in the NBG state with different depths,
$d$. In Figs.~\ref{figw}(a,b,c) the last processes are related to this new
 NBG state. Note that when we do not have any compression in the
system both $\Delta F$ and $W$ are zero as it is clear from the flat
regions during process `II' in Figs.~\ref{figw} (a,b). In order to
be sure about the stability of the NBG at different values of
strain, we performed extra long simulations for the process `II' up
to nanoseconds. Figure~\ref{figbowl} shows two final snapshots of
two long time simulations during process `II'. In this figure $a=10$
nm and total applied strains are 1.2$\%$ and $6\%$ for Fig. 3(a) and
Fig. 3(b), respectively. As we see, a  larger strain results in a
deeper \emph{nano-bowl}. These NBGs are stable and their concave
shape survives independently of the considered temperature and
strains. Therefore they keep their shapes and are static. The
negative (positive) value for $\Delta F$ at $T$=300~K ($T$=50~K)
during the evolution S$\rightarrow$II indicates that the NBG at this
temperature is more (less) stable than the starting state `S'. Note
that  gravity and other external fields have been ignored and the
bowl-like shape and the hump shape occur random with the same
probability, and we call both shapes NBG. This concave shape should
be observable experimentally using scattering experiments.

When comparing the difference of the free energy between `M' and
NBG, using the values in Table~\ref{table1}, gives $\Delta
F(M\rightarrow II)$=3.2965~eV for $T$=300~K which indicates that the
real optimum equilibrium state is `M'. Here, in all cases we see the
validity of the inequality Eq.~(\ref{Wpositive}).
 Note that the differences in the free energy per atom are
very small, i.e. $\leq10^{-4}$eV.

\begin{table}
\begin{tabular}{|c|c|c|c|c|c|}
 \hline
 $T$(K)& $\langle W\rangle(S\rightarrow M)$&$F_M-F_S$&$E_M-E_S$&$R_{M}$&$\sigma_w$\\
 \hline
 300&-3.9514 (eV)&-3.9940(eV)&$\sim$ 23(eV)&$a-0.3\delta$&0.047(eV)\\
50&-0.5164(eV)&-0.5688(eV)&$\sim$ 2(eV)&$a-0.11\delta$&0.021(eV)\\
     \hline
\end{tabular}
  \centering
\caption{Total work done on the system $\langle W\rangle$, difference of the free
energy (Eq.~(\ref{jeq})), difference of the total energy, the
equilibrium radius of points `M' and variance in work distribution
at points `M' for the system with initial radius $a=10$~nm~at
$T$=300~K and 50~K (see Figs.~\ref{figw}) for
$\epsilon=1.2\%$.}\label{table1}
\end{table}

\section{Conclusions}
In summary, when applying an increasing radial force on the boundary
of circular monolayer graphene, first a linear response regime
appears which enables us to estimate Young's modulus and
pre-stresses, which are found to be in agreement with experiments.
When continuing compression, at a critical radial force, graphene
starts to buckle. The computed critical buckling strains are larger
than those found from elasticity theory. These states are new
thermodynamically buckled stable states which we called
\emph{nano-bowl}. In order to investigate the thermodynamical
stability of \emph{nano-bowls}, we used a non-equilibrium
computational method based on the Jarzynski identity. This identity
enables us to calculate the difference in the free energy between
the initial non-compressed state and the buckled states. Also the
optimum radius of circular graphene, where there is no boundary
stress, can be estimated by looking at the minimum in the free
energy curve. At room temperature the \emph{nano-bowl} is more
stable than the initial non-compressed state which was pre-stressed.
Free energy calculations based on the Jarzynski~\cite{jar} equality
can open a new approach to study various thermomechanical properties
of compressed graphene.

\section{Acknowledgment}
 This work was supported by the Flemish Science Foundation (FWO-Vl) and the Belgian Science Policy~(IAP).

\end{document}